\documentclass[twocolumn,showpacs,aps,preprintnumbers,amsmath,amssymb, floatfix,altaffilletter]{revtex4}

\usepackage{graphicx}
\usepackage{bm}


\begin{document}

\title{Femtosecond x-ray absorption spectroscopy of spin and orbital angular momentum in photoexcited Ni films during ultrafast demagnetization}

\author{C. Stamm} 
\email{christian.stamm@helmholtz-berlin.de}
\author{N. Pontius}
\author{T. Kachel}
\author{M. Wietstruk}
\author{H. A. D\"urr}
\altaffiliation{also at: SLAC National Accelerator Laboratory, Menlo Park, 
CA 94025, USA }
\email{hdurr@slac.stanford.edu}
\affiliation{BESSY II, Helmholtz-Zentrum Berlin f\"ur Materialien und Energie GmbH, Albert-Einstein-Strasse 15, 12489 Berlin, Germany}

\date{\today}

\begin{abstract}

We follow for the first time the evolution of the spin and orbital angular momentum of a thin Ni film during ultrafast demagnetization, by means of x-ray magnetic circular dichroism. 
Both components decrease with a $130 \pm 40$~fs time constant upon excitation with a femtosecond laser pulse. 
Additional x-ray absorption measurements reveal an increase in the spin-orbit interaction by 
$6 \pm 2$~\% during this process. 
This is the experimental observation of a transient change in spin-orbit interaction during ultrafast demagnetization. 
\end{abstract}

\maketitle

\newcommand{\so}{\langle \textbf{S} \cdot \textbf{L} \rangle}

The manipulation of long-range ferromagnetic order by a femtosecond (fs) laser pulse is known for more than a decade \cite{BeaurepairePRL76}.
In spite of its implications for overcoming speed limits in conventional magnetic data storage \cite{TudosaNature428},
the underlying microscopic mechanism remains unknown. Experimental evidence for femtosecond demagnetization of 3$d$ transition metal systems such as ferromagnetic Ni \cite{BeaurepairePRL76,HohlfeldPRL78,SchollPRL79,RhiePRL90,StammNM6} or Co \cite{CinchettiPRL97} were interpreted in terms of spin flip scattering processes during laser-excited hot electron relaxation \cite{KoopmansPRL95}. 
\textit{Ab initio} calculations for Ni seem to confirm the feasibility of this
scenario for spin angular momentum transfer to the lattice \cite{SteiaufPRB79}. 
However, it was argued that the combination of spin-orbit coupling and a strong laser field could lead to a novel excited state promoting ultrafast demagnetization \cite{ZhangPRL85}. 
Despite many attempts, only very recently experimental evidence was reported which could support such a scenario \cite{BigotNatPhys5}. 
However, the validity of the employed magneto-optic experimental probe has been debated in the literature \cite{KoopmansPRL85,ZhangNatPhys5}.
It is, therefore, of paramount importance to characterize the possible laser-induced transient state by independent experimental tools.
Such studies will also have significant impact on the recently reported all-optical magnetic switching in transition metal, rare-earth compounds \cite{VahaplarPRL103}.

Here we report the direct observation of a transient increase in spin-orbit coupling during laser-induced ultrafast demagnetization of a thin ferromagnetic Ni film. 
This is correlated with the decay of spin, $S$, and orbital, $L$, momentum. 
We use sum rule analysis of the time-resolved x-ray absorption spectroscopy (XAS) and x-ray magnetic circular dichroism (XMCD) signals. 
These techniques do not suffer from the state-bleaching effects observed in optical spectroscopy \cite{KoopmansPRL85}, since initial and final states are different for optical excitation and x-ray probe \cite{StammNM6,CarvaEPL86}. 
The fs laser pulse causes a decrease in both $S$ and $L$ from their equilibrium value down to $\approx 10$~\%, with a time constant of 130~fs as detected by XMCD. 
During this process, a transient increase in $ \langle\, \textbf{S} \cdot \textbf{L} \, \rangle $, the expectation value of the spin-orbit operator, by 6~\% is observed by XAS. The observed temporal behavior allows us to relate this to the driving force for ultrafast demagnetization.

The experiments were performed at the BESSY II femtoslicing source \cite{HolldackPRSTAB8,KhanPRL97}, 
which produces photons of freely selectable polarization in the soft x-ray range with duration of 100~fs. 
The newly designed beamline UE56/1-ZPM is optimized for the very low photon flux from the fs x-ray source. 
It produces a 20 times higher x-ray intensity compared to the standard beamline used in our first fs laser pump -- fs x-ray probe experiments \cite{StammNM6}. 
This is achieved by using a single optical element, a Bragg-Fresnel reflection zone plate, for focusing and energy dispersing the x-ray beam \cite{MichetteOC245}. 
X-rays tuned to the $L_{2}$ and $L_{3}$ absorption edges were transmitted through a thin film sample, promoting the  2$p_{1/2}$ and 2$p_{3/2}$ electrons, respectively, into the 3$d$ valence shell. 
XMCD was measured using circularly polarized x-rays, with an applied magnetic field of $\mu_0H=0.24$ T that was reversed at each point of the scan, oriented collinear to the x-ray beam.
As XMCD measures the projection of the magnetic moments onto the x-ray propagation direction, a $35^\circ$ angle of incidence on the in-plane magnetized sample was used.
Additionally, XAS data were taken with linearly polarized x-rays in normal incidence and without applying a magnetic field.
With these experimental conditions we exclude any influence of the sample magnetization on the XAS data.
The samples consisted of 17~nm and 35~nm thick Ni films, evaporated \textit{in situ} under ultrahigh vacuum conditions onto a free-standing Al foil of 500~nm thickness.
Characterization of the structural and magnetic properties was carried out by x-ray absorption and hysteresis loop measurements using XMCD.
The laser setup consists of an amplified Ti:Sapphire laser capable of a pulse energy of 2~mJ at a repetition rate of 3~kHz, 780~nm wavelength, and pulse length of 60~fs. 
The laser beam is split, and 85~\% of the initial energy is used for the generation of the ultrashort x-ray pulse in the storage ring. 
The remaining 15~\% are guided onto the sample as pump pulse, after passing a chopper blocking every other laser pulse. 
This allows us to simultaneously record the x-ray absorption of the pumped and the non-pumped sample in the same measurement. 
The laser spot diameter on the sample was $\approx$\,1~mm, and the x-ray probe size $\approx$\,0.2~mm.

\begin{figure}
\includegraphics[width=80mm]{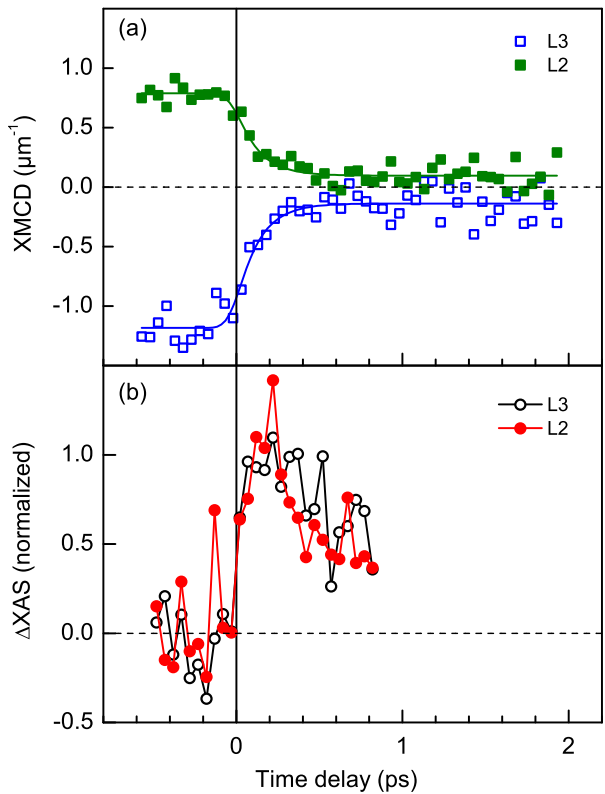}
\caption[]{(Color online)
Time-resolved XMCD and XAS measurements on a 17~nm Ni film. 
In (a), the pump-probe delay scans of the XMCD signals are plotted, measured at the $L_3$ and $L_2$ absorption edges, with a pump fluence of $12~\mathrm{mJ/cm^2}$.
The lines are fits with an exponential decay with 130~fs time constant, convolved with a Gaussian with 130~fs FWHM. 
In (b), the laser-induced changes in absorption on the leading $L_3$ and $L_2$ edges are plotted, normalized to the same amplitude. The photon energy was set to maxima found in Fig.~\ref{XAS}(b). 
Pump fluence was $16~\mathrm{mJ/cm^2}$.
} 
\label{delay}
\end{figure}

In Fig.~\ref{delay} we show the temporal evolution of the XMCD and XAS data after exciting the Ni film with a fs laser pulse.
Starting from a constant dichroism before time zero in Fig.~\ref{delay}(a), the XMCD signal measured at the $L_{2,3}$ absorption edges is reduced to  10~\% of its initial value for time delays longer than 1~ps. 
Fitting with an exponential decay, convoluted by a Gaussian of 130~fs [full width at half maximum (FWHM)] representing the experimental resolution, results in the same time constant of $130 \pm 40$~fs for the XMCD dynamics at the $L_2$ and $L_3$ edges. 
This complete dichroism measurement allows us in the following to determine the evolution of spin and orbital angular momentum in the fs time range.

In addition to the XMCD, we find in Fig.~\ref{delay}(b) that also the absorption of linearly polarized x-rays
is influenced by the intense laser pump pulse. 
The effect is more subtle than the one observed with XMCD. 
It has first been observed on the $L_3$ edge \cite{StammNM6} and the $L_3$ satellite \cite{KachelPRB80}, the latter however only in thermal equilibrium 100~ps after laser heating. 
As we will demonstrate in the following, a comparison of the dynamic XAS measurements of the two absorption edges, $L_2$ and $L_3$, on the fs time scale shown in Fig.~\ref{delay}(b) reveals important new information about the valence electron system in the laser-excited sample.
Note that the two difference curves in Fig.~\ref{delay}(b) exhibit the same temporal behavior within the experimental accuracy.
This behavior is, however, distinctly different from the XMCD signals in Fig.~\ref{delay}(a): the XAS in Fig.~\ref{delay}(b) peaks at delay times while the XMCD in Fig.~\ref{delay}(a) is still in the progress of quenching. We will show in the following that this reflects the driving force behind ultrafast demagnetization.

\begin{figure}
\includegraphics[width=80mm]{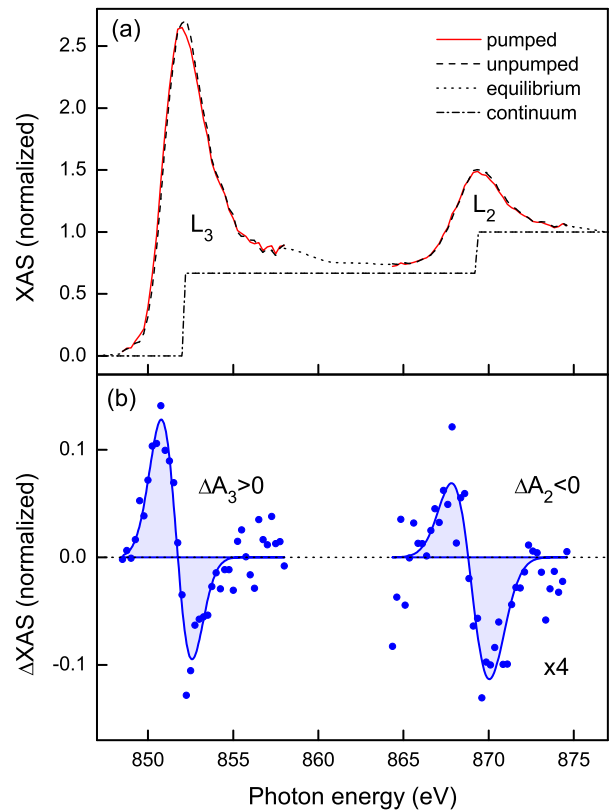}
\caption[]{(Color online) 
XAS of a laser-excited 35~nm Ni film, measured at 200~fs pump-probe delay with linearly polarized x-rays. 
The absorption with and without laser pumping is drawn in (a) as continuous and dashed line, respectively. 
The dotted line represents an equilibrium absorption spectrum, measured independently without time resolution.
A double step function at $L_3$ and $L_2$ with amplitude ratio $2:1$ is drawn as dashed-dotted line.
All data are normalized to an edge jump of 1. 
In (b), the difference between the pumped and the unpumped absorption is plotted.
The shaded areas are guide to the eyes to highlight the difference of $\varDelta A_3$ and $\varDelta A_2$.
Energy resolution of the spectra was $1.5$~eV, and laser pump fluence was $\approx 12~\mathrm{mJ/cm^2}$.
} 
\label{XAS}
\end{figure}

To obtain a better understanding of the laser-induced XAS changes in Fig.~\ref{delay}(b),
we next look at the absorption spectrum of the Ni film 200~fs after laser excitation and compare it to the spectrum of the unpumped sample, which was measured in parallel. 
The respective absorption edges shown in Fig.~\ref{XAS}(a) exhibit a small shift to lower photon energies for the laser-excited film.
These shifts of the XAS threshold are better visible in the difference spectra plotted in Fig.~\ref{XAS}(b). 
They are discernable as the derivative-like line shapes occurring at both absorption edges. 
The fact that both absorption edges display shifts of identical sign, although of different magnitude, demonstrates the purely electronic origin of this effect. 
Note that any magnetic signatures should display opposite signs since spins and orbits are coupled (anti)parallel at the respective spin-orbit split absorption edges. 
We previously assigned the shift in absorption threshold to a transient 2$p$ core-level shift induced by a valence-band narrowing, i.e., an increased electron localization \cite{StammNM6,KachelPRB80}.  
Recent \textit{ab initio} calculations of laser-excited Ni films \cite{CarvaEPL86} reproduced our experimental findings, demonstrating that also repopulation of electronic levels plays a role.

\begin{figure}
\includegraphics[width=80mm]{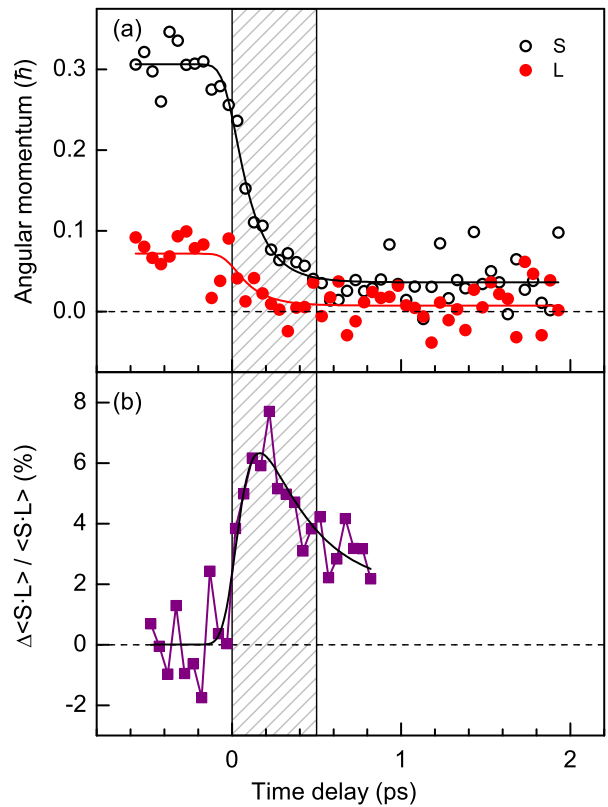}
\caption[]{(Color online) 
(a) Time-resolved spin and orbital angular momentum, obtained from XMCD data in Fig.~\ref{delay}(a) using sum rules.
The lines are exponential fits with a 130~fs time constant. 
(b) Time-resolved changes in the spin-orbit interaction, $\so$.
The shaded area marks the time during which the demagnetization process is active.
The line in (b) is a fit with two exponentials with decay times of 100~fs (rise) and 300~fs (relaxation, for details see Ref.~\cite{StammNM6}).
All fits are convolved with a Gaussian of 130~fs FWHM.
} 
\label{eval}
\end{figure}

Integrating over the respective absorption edges in Fig.~\ref{XAS}(b) reveals a further effect. 
This is illustrated by the two Gaussians with opposite amplitudes drawn as guide to the eyes along the $\varDelta A_{2,3}$ data of the individual absorption edges. 
The integral is positive at $L_3$ but negative at $L_2$, indicating the magnetic origin of these total intensity changes.
It also follows that the branching ratio\cite{LaanPRL60}, defined as $B = A_3 / (A_3 + A_2)$, is enhanced in the laser excited sample.
Absorption into continuum states is accounted for by a double step function [see Fig.~\ref{XAS} (a)] and is subtracted in our evaluation of the $A_{2,3}$ values. 
We find $B=0.75$ in equilibrium, and a relative change in $\varDelta B / B = 0.63 \pm 0.22$~\% for the excited state.
At the same time, the total absorption summed over both edges $A_2 + A_3$ is unchanged within the experimental error: the relative XAS change is $(\varDelta A_2 + \varDelta A_3) /(A_2 + A_3)= 0.3 \pm 0.6$~\%.
As a direct consequence, the number of holes in the $3d$ valence shell, $n_h$, remains constant during the excitation process \cite{TholePRB38}.
This fact will be used later in the sum rule analysis.
A linear relationship links the branching ratio to the expectation value of the angular part of the $3d$ valence band spin-orbit coupling operator $\textbf{S} \cdot \textbf{L}$ as \cite{TholePRB38}:
\begin{equation}	 
	B = B_0 +  \langle\, \textbf{S} \cdot \textbf{L} \, \rangle 
\end{equation}
Using the branching ratio sum rule and the so-called statistical value of $B_0 = 2/3$
\cite{LaanPRL60}, we can determine the relative change 
$\varDelta\so / \so$ to $6 \pm 2$~\%.
The temporal evolution of $\varDelta\so$ is obtained by scaling the static values of $\varDelta A_{2,3}$ taken from Fig.~\ref{XAS}(b) with the dynamic XAS curves in Fig.~\ref{delay}(b). 
The result is plotted in Fig.~\ref{eval}(b).

Now we turn to the evaluation of the time-resolved spin and orbital angular momentum through sum rule analysis \cite{TholePRL68,CarraPRL70}.
Therein, $S$ and $L$ are given by linear combinations of the dichroism, $D_{2,3}$, at the $L_{2,3}$ edges:
\begin{eqnarray} 
S &=& - \frac{1}{2}\frac{D_3 - 2 D_2}{A_3 + A_2} \cdot  \hbar\, n_h
 \\ \nonumber \\
L &=& - \frac{2}{3}\, \frac{D_3 + D_2}{A_3 + A_2} \cdot  \hbar\, n_h
\end{eqnarray}
Here $D_{2,3}$ denote the difference in absorption for parallel and antiparallel orientation of photon helicity and applied magnetic field, integrated across the absorption edge.
A dichroism spectrum measured with the sample in equilibrium (not shown here), using $n_h = 1.66$, was used to determine the quantitative values of $S$ and $L$ before time-zero.
To obtain the time-resolved evolution of $S$ and $L$ plotted in Fig.~\ref{eval}(a), we used the transient XMCD data in Fig.~\ref{delay}(a). These were acquired with an energy resolution of 5~eV, which essentially integrates the dichroism across the absorption edge.
Here we neglected the so-called magnetic dipolar term, a correction that typically influences the spin value on a 10~\% level \cite{ChenPRL75}. It was shown that the magnetic dipole term averages to zero in angle averaged measurements or on polycrystalline samples as used here \cite{StohrPRL75,DurrPRB54}.

Previous measurements already showed that the total angular momentum $J = S + L$ decreases on the fs time scale indicating an ultrafast angular momentum transfer to the lattice \cite{StammNM6}. Here we separate $S$ and $L$ for the first time. A very clear drop of the spin momentum is observed with the $130 \pm 40$ fs time constant as determined from fitting the XMCD curves in Fig.~\ref{delay}(a). 
Also for the orbital momentum a fast drop induced by the laser-pulse is visible. 
Due to the large experimental error, we refrain from determining an individual time constant for $L$.
However, the data for $L$ are well reproduced by scaling the curve found for $S$.

The obtained values $S$ and $L$ represent the projection of the corresponding vector quantities onto the x-ray incidence direction \cite{TholePRL68,CarraPRL70,ChenPRL75,StohrPRL75,DurrPRB54}. 
They are also averaged over the probing volume and, thus, measure the long-range ferromagnetic order. 
The almost complete loss of this ordering coincides with another process: 
the transient increase in the spin-orbit interaction, $\varDelta\so$, plotted in Fig.~\ref{eval}(b).
Both processes take place during a time interval of about 500~fs, initiated by the laser pump pulse.
During this time interval, the Ni film looses 90~\% of its magnetization. 
The associated total angular momentum, $J$,
needs to be transferred to the lattice due to conservation laws. 
Exactly within this time window, we observe a peak in $\varDelta\so$. 
As will be described in the following, this observation provides access to the microscopic driving force for demagnetization.

In the presence of spin-orbit coupling $S$ and $L$ are no longer good quantum numbers, i.e., spin angular momentum can be transferred to the electron orbits and vice versa. This, however, conserves $J$ unless there is coupling to the lattice \cite{KoopmansPRL95,CarpenePRB78,ZhangPRL101}. 
Our data in Fig.~\ref{eval} provide new boundary conditions for identifying this latter process. 
It has been argued that $S$ must be temporarily transferred to $L$ before the latter is quenched via crystal-field excitations \cite{CarpenePRB78}. 
Figure~\ref{eval}(a) suggests that this process is very effective, since $S$ and $L$ decay simultaneously.

A conceptionally very similar picture is Elliott-Yafet spin-flip scattering, where spin-polarized electrons are scattered into regions of the Brillouin zone with a strong spin-orbit coupling, the so-called hot spots \cite{KoopmansPRL95,SteiaufPRB79,CinchettiPRL97}. 
In hot spots, spin-up and spin-down bands in ferromagnets are mixed due to spin-orbit coupling \cite{SteiaufPRB79}. 
The characteristic feature of hot spots is that they contribute significantly to $\so$ but not to $S$. 
A transient accumulation of charge carriers in hot spots should therefore lead to a transient increase in $\so$. 
This is exactly what we observe in Fig.~\ref{eval}(b). 
The maximum of $\varDelta\so$ coincides with the maximum slope of $S$ indicating that spins are flipped by being scattered into regions with increased $\so$. The corresponding change in spin angular momentum is effectively transferred to the lattice in this process without increasing $L$ \cite{CarpenePRB78}. 
Finally the electrons are scattered away from the hot spots again, however, without flipping spins back. 
This reduces $\so$ but leaves $S$ quenched at longer time delays in Fig. 3. A possible mechanism for the latter is that spin-flips decay into magnons which leave the sample demagnetized \cite{CarpenePRB78,CinchettiPRL97}.

We finally mention that two different scattering processes can lead to spin flip scattering.
Spins may be coherently excited during the driving laser field \cite{ZhangPRL101,BigotNatPhys5}. 
Within the two-state model of Ref.~\onlinecite{ZhangPRL101}, this may be explained by a coherent (de)population of states with different spin-orbit coupling. 
This fact alone would neither change $\so$ nor the sample magnetization. 
However, the response of the many-electron system leads to a dephasing of the coherent laser excitations \cite{ZhangPRL101}. 
Presently, the temporal resolution is insufficient to conclusively establish the importance of such coherent excitations. 
However, the temporal correspondence of the increase in $\so$ with the electronic thermalization, as well as the observation of magnetization dynamics after the driving laser pulse has passed, indicates the importance of a second, incoherent process. 
In this case, spins can be scattered by other (laser-excited) electrons into hot spots \cite{KoopmansPRL95,CinchettiPRL97}. 
This was clearly observed for Co \cite{CinchettiPRL97}, where demagnetization also occurs after the driving laser pulse is over.

Summarizing, upon pumping with a fs laser pulse a thin Ni film reacts with a simultaneous decrease in spin and orbital angular momentum with a $130 \pm 40$~fs time constant. 
During this demagnetization process, the spin-orbit coupling is altered such that $\so$ increases by 6~$\pm$~2~\% mainly while the pump laser field is on.
This is the first direct experimental proof of the importance of the spin-orbit interaction during ultrafast demagnetization.

We thank T. Quast, R. Mitzner, and K. Holldack for their help during the measurements, 
and A. Erko and A. Firsov for their expertise in designing and fabricating the zone plate optics.




\end{document}